# Actinium-225 Production with an Electron Accelerator

W. T. Diamond (1) and C. K. Ross (2)

January 2021

(1) Retired, Atomic Energy of Canada limited, Chalk River Laboratories, Chalk River, Ontario,

 Canada K0J 1J0, diamond_w45@yahoo.ca

(2) Retired, National Research Council, Ottawa, Ontario, Canada K1A 0R6,
carlkross@gmail.com

**Abstract**

There has been growing clinical evidence of the value of targeted alpha therapy for treatment of several cancers. The work has been slowed by the lack of availability of the key alpha emitting isotopes, especially Ac-225. Until this time, most of the supply has been from three Th-229 generators that are milked to produce hundreds of mCi of Ac-225 every month. There has been a growing effort to produce new sources of Ac-225 from several different accelerator-based routes. It can be produced with medical-isotope cyclotrons with a proton energy of at least 16 MeV using the reaction Ra-226(p,2n)Ac-225. It can also be produced by using high-energy protons (150 to 800 MeV) for spallation of a thorium target. Significant experimental work has been applied to both processes. It can also be produced by the photonuclear reaction, Ra-226(γ,n)Ra-225. The Ra-225 decays via beta decay to Ac-225 with a half life of 14.9 days. The photons are produced by an intense beam of electrons with an energy about 25 to 30 MeV. This paper will provide a technical description of radium targets and a target chamber that would be capable of producing a yield of four curies of Ra-225 from a 10-day irradiation of one gram of radium segmented into two to four separate encapsulated targets, at a beam power of 20 kW. These targets could be milked at least three times, yielding nearly four curies of Ac-225. There is also a description of a method to reduce production of Ac-227 to values less than a few parts per million of the yield of Ac-225. The Monte Carlo code Fluka has been used to model the yields of Ra-225 and support the design concept to reduce the production of Ac-227. It has also been used to model the experimental results by Maslov et al. [https://doi.org/10.1134/S1066362206020184] to provide reasonable confidence in the cross-section value used by the code.

1. INTRODUCTION

Targeted alpha therapy (TAT) has been demonstrated to be a promising approach for the treatment of cancer. There have been recent trials that show the positive effects of using $^{225}$Ac and one of its daughter products, $^{213}$Bi [1, 2]. Most of this work has been done using three $^{229}$Th sources located in the USA, Russia and Germany [3]. $^{229}$Th decays via alpha decay ($t_{1/2}$ = 7340 y) to $^{225}$Ac that can be milked from the thorium source multiple times per year. These produce only a limited supply $^{225}$Ac [4], insufficient for development of broader uses of the various therapies. A recent joint IRC-IAEA workshop [3], "Supply of Actinium-225" produced a good summary of the recent clinical results and efforts to produce greater supplies of $^{225}$Ac. Three different production routes using accelerators were reviewed:

> Spallation of $^{232}$Th using high-energy protons;
>
> $^{226}$Ra(p,2n)$^{225}$Ac using a radium target and typical medical-isotope cyclotrons of proton energies above about 16 MeV;
>
> $^{226}$Ra(γ,n)$^{225}$Ra that beta decays ($t_{1/2}$ = 14.9 d) to $^{225}$Ac. The photons can be produced by electrons of about 25 to 30 MeV.

There is significant ongoing research on the first two processes [5, 6] and some effort on using an electron linac [7, 8]. This paper will provide a detailed overview of the use of an electron accelerator to produce about four Ci of $^{225}$Ra from a 10-day irradiation of a one-gram radium target at 25 MeV and 20 kW of electron beam power. If milked three times over a 45-day period, this target would yield about four Ci of $^{225}$Ac.

2. ELECTRON ACCELERATOR PRODUCTION OF $^{225}$Ac

**2.1 Use of Radium as a Target Material**

Radium as a target material presents serious challenges. It is a highly radiotoxic element and the heaviest alkaline metal from Group 2 of the periodic table. As such it is a highly reactive metal that also has poor physical, chemical and thermal properties for use as a target material. Production of reasonably high yields of $^{225}$Ac requires the use of the order of 100's of mg to about a maximum of one gram [9] of radium. One curie (37 GBq) was originally defined as the quantity or mass of radium emanation in equilibrium with one gram of radium. Therefore, practical radium targets for high-yield production of $^{225}$Ac produce the order of 10 to 37 x 10$^9$ alpha decays per second that are in secular equilibrium with four more alpha decays. These alphas become helium atoms dispersed throughout the target material. During production of multi-curie yields of $^{225}$Ra over irradiation periods of ten to fifteen days, there are many more atoms of helium produced via the decays of the $^{225}$Ac isotope and the other co-produced isotopes, especially $^{224}$Ra. The first decay product of $^{226}$Ra is $^{222}$Rn that is an alpha emitter and a gas at room temperature. Most stages of material handling will need appropriate filtration to capture the radon gas and hold it until it decays ($t_{1/2} = 3.862$ d).

The alpha decay is also accompanied by significant beta and gamma decay. Table 1 shows the expected gamma-radiation field from one curie of $^{226}$Ra and from the other products produced during the photon irradiation of a radium target. The calculations were done with the shielding code MicroShield [10] and assume secular equilibrium with all decay products. Some of the isotopes will not be in secular equilibrium at EOB (End of Bombardment) and may produce significantly lower gamma radiation fields than calculated. However, $^{224}$Ra that is co-produced in large quantities during 10- to 15-day irradiation via an electron accelerator, has a half life of 3.6 d and will reach near secular equilibrium at EOB. It produces highly penetrating gamma radiation that requires additional shielding.

The target designs proposed in Section 2.4 use from two to four separate hermetically encapsulated targets. This would reduce the maximum radiation field for each target to ½ to as low as ¼ except for the time that they are assembled in the target holder before or after irradiation. Most parts of target handling would be done using remote handling and some of the features described in Section 2.4 are designed to help enable that.

Table 1. Gamma radiation fields at one meter from a one-curie (37 GBq) source of $^{226}$Ra, $^{225}$Ra, $^{224}$Ra and $^{225}$Ac. The Microshield [10] calculations assume secular equilibrium of all decay products. Note that 10 cm of Pb would provide adequate shielding for a radiation worker who spends a few hours per year working with the targets.

| Isotope | Half Life | Unshielded mR/h | 5 cm of Pb mR/h | 7.5 cm of Pb mR/h | 10 cm of Pb mR/h |
|---|---|---|---|---|---|
| Ra-226 | 1600 y | 906 | 27 | 6.3 | 1.5 |
| Ra-225 | 14.9 d | 108 | 0.06 | | |
| Ra-224 | 3.6 d | 755 | 38 | 11 | 3 |
| Ac-225 | 9.9 d | 92 | 0.06 | | |

The thermal properties that are of significance in the use of radium as a target material are:

Thermal conductivity of 19 W/m-K;

Melting point of 690 °C;

Specific heat of 0.12 J/g-K.

These poor thermal properties limit the maximum power density that can be applied to the target.

### 2.2    Photonuclear Reactions on Solid Targets

High-energy photons (above about 6 MeV) can produce photonuclear reactions with nuclei. A suitable photon spectrum can be produced by the bremsstrahlung radiation arising from the interaction of electrons with either the target material or with a high-Z target ahead of the isotope target, called a converter target. The radiation length, $X_o$, [11] is the distance that a high-energy electron travels so that its energy is reduced due to bremsstrahlung by an average factor of 1/e.

For reference, the values of $X_o$ for radium is 6.15 g/cm$^2$ or 12.3 mm and for tungsten (tantalum) are 6.76 (6.82) g/cm$^2$ or 3.5 (4.1) mm [11]. For low-Z material such as aluminum or cooling water the values of $X_o$ can reach many centimetres and these materials produce significantly less attenuation of the bremsstrahlung beam compared to target materials when they are used as part of a target or cooling system.

The bremsstrahlung produced by the interaction of an electron beam with a target is largely forward focused. The approximate angle at which the energy flux is reduced by 50% is given as [11]:

$$E_o \Theta_{1/2} \sim 100 \qquad (1)$$

where $E_o$ is the energy of the electrons in MeV and $\Theta_{1/2}$ is the 1/2 -angle given in degrees. For electron energies above 25 MeV, $\Theta_{1/2}$ is smaller than 4 degrees.

For high-Z targets the dominant reactions are ($\gamma$,n) and ($\gamma$,2n) but the ($\gamma$,n) reaction is generally the more intense reaction. A thick target is about five radiation lengths and a target thickness of one (two) $X_o$ produces about 34% (56%) of the thick target yield [11]. For high-value targets such as radium the amount used will be significantly less than one radiation length and it will be necessary to use a converter target ahead of the isotope target to optimize the yield. If an optimized (about 0.5 to 0.75 $X_o$) converter target is used the photonuclear yield of the isotope-production target is approximately linear with its length compared to 1 $X_o$, i.e., 34% of thick target yield times the thickness in radiation lengths [11].

The use of thin targets leads to a significant reduction in yield compared to the use of a thick target. This can be partially compensated using higher beam power than can be used with a positive ion beam. The target can also be segmented into several sections that are individually encapsulated with a material such as aluminum. One-half mm of aluminum on both sides of a radium target produces little reduction in the bremsstrahlung intensity, even when multiple segments are used, but provides excellent protection against target rupture during irradiation and handling. Aluminum is a good choice for target cladding because it has been used in high-powered electron beams with little degradation from such processes as oxidation by the radiolysis products of water. It also has little residual activity at the EOB for electron energies below 30 MeV [12].

Radium salts such as radium chloride can be used as target material but the extra material introduced by using a salt has some impact on the yield from the reaction. The compounding material such as chlorine needs to be evaluated for the potential reactions that may make handling the irradiated target more challenging. Radium nitrate, radium carbonate and radium sulphate only produce short half-life products such as $^{13}$N ($t_{1/2}$ = 10 min) and $^{15}$O ($t_{1/2}$ = 122 s) for radium nitrate and $^{11}$C ($t_{1/2}$ = 20.4 min) and $^{15}$O ($t_{1/2}$ = 122 s) for radium carbonate.

A review of barium compounds (barium is the nearest alkaline metal to radium from Group 2 of the periodic table) shows that a compound such as BaB$_6$ (barium hexaboride) is a very stable compound that has superior thermal properties to barium metal (melting point of 2200$^o$C and thermal conductivity of about twice that of radium metal). It can be produced in small quantities [13]. The only significant radioactive product produced by irradiating boron is a low yield of $^{10}$Be with a half-life of 1.5 million years. If a method can be found to extract the $^{225}$Ac from radium hexaboride then this target material might be capable of use at target powers above 20 kW.

## 2.3 Yield Estimate

As stated by Martin et al. [14], the rate of radionuclide production is given by

$$RR = N_T \int^{E_o} \sigma(E)\varphi(E,E_o)dE \qquad (2)$$

where

RR is the reaction rate, or the number of nuclei per unit volume formed per second

$N_T$ is the number of target atoms per cm$^3$

$\varphi(E,E_o)$ is the incident particle flux per cm$^2$ per second, and

σ is the cross section in cm$^2$

The activity at EOB is given by:

$$A = RR(1 – \exp(-\lambda t)) \qquad (3)$$

where t is the irradiation time and λ is the decay constant.

Figure 1 shows a Monte Carlo calculation of the bremsstrahlung spectrum for 25 MeV electrons incident on a 3-mm tantalum target typical of the thickness of a converter target. This figure shows the number of bremsstrahlung photons averaged over the second radium target (see Figure 2) per cm$^2$ for one mA of 25 MeV electrons. The spectrum was produced by the Monte Carlo code Fluka [15]. A second Fluka calculation was done to compare the bremsstrahlung spectrum at 30 MeV produced by a 3-mm tungsten converter to the spectrum given by Martin et al. [14]. There was close agreement.

Figure 1 also includes the TENDL [12] (γ,n) and (γ,2n) cross sections for a $^{226}$Ra nucleus, referenced to the right-hand y-axis. There are no measured cross sections for comparison. The (γ,f) and any of the other possible reactions have a cross section that is less than a fraction of one percent of the (γ,xn) cross sections.

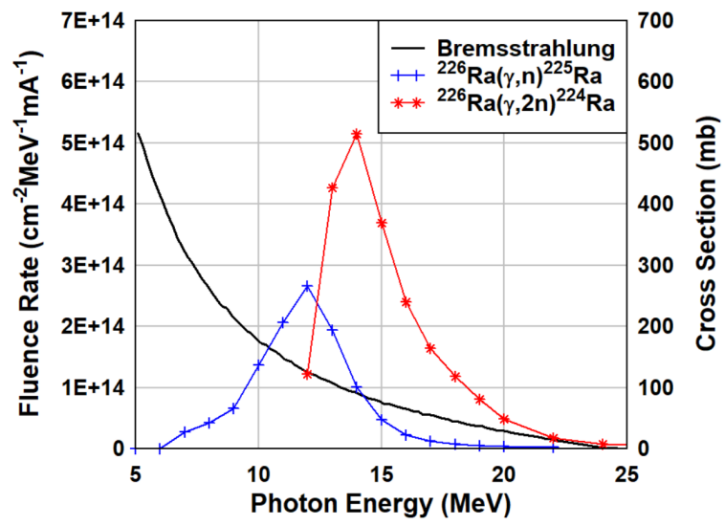

Figure 1. TENDL [12] cross sections of $^{226}$Ra(γ,n)$^{225}$Ra and $^{226}$Ra(γ,2n)$^{224}$Ra and the bremsstrahlung spectrum produced by 25-MeV electrons on a 3-mm tantalum converter.

### 2.3.1 $^{225}$Ra Yield Estimate Based on Photonuclear Systematics

An initial yield estimate can be obtained by examining the systematics of the photonuclear production for nuclei of similar mass to the radium target. Figure 32 in reference 11 shows the neutron yield for a natural lead target as a function of electron-beam energy for five target thicknesses. For a 1 $X_o$ natural lead target the neutron yield is about 3.0 x $10^{11}$ n/s/kW for 25 MeV electrons and the (γ,n) reaction accounts for about 80% of this yield [16] for each of the three stable isotopes ($^{206}$Pb, $^{207}$Pb and $^{208}$Pb) that account for over 98% of the abundance of natural lead. This produces a yield of about 2.4 x $10^{11}$ n/s/kW for the (γ,n) reaction on natural lead at 25 MeV. The TENDL [12] (γ,n) cross section of $^{226}$Ra is about 40% of the (γ,n) cross section of $^{208}$Pb. This would lead to an initial estimate of 9.6 x $10^{10}$ n/s/kW from a 1 $X_o$ target of radium irradiated at 25 MeV from the $^{226}$Ra(γ,n)$^{225}$Ra reaction. A one-gram target of one-cm diameter is 2.56 mm thick or 2.56/12.3 = 0.21 $X_o$ thick. With a near-optimized converter target ahead of the radium target the yield should be at least as high as this fraction times the yield for a 1 $X_o$ target. Therefore, the initial estimate is:

9.6 x $10^{10}$ x 0.21 = 2.0 x $10^{10}$ n/s/kW at 25 MeV from the $^{226}$Ra(γ,n)$^{225}$Ra reaction

This is equivalent to the reaction rate or the saturated yield. The estimated activity after 10-day irradiation at 25 MeV is:

7.4 GBq (EOB) or 0.20 Ci (EOB) per kW of electron beam power.

A 20-kW irradiation would produce:

4.0 Ci (EOB) for a 10-day irradiation.

## 2.3.2 Monte Carlo Calculations of $^{225}$Ra Yields

The Monte Carlo code Fluka [15] has been used to provide estimates of the yield for several configurations of a converter target and a radium target and has been used to help optimize the target design. The initial calculation was based upon the following basic setup that should produce about optimal results for a target mass limited to one gram of radium:

25 MeV pencil beam, 5 mm in diameter;
Tantalum bremsstrahlung target, 10 mm in diameter, 3 mm thick;
2 mm vacuum gap to Ra target;
$^{226}$Ra target, 10 mm in diameter, 2.56 mm thick, for a mass of 1 g.

The result was an activity of 5.36 Ci (EOB) of $^{225}$Ra for a 20-kW irradiation at 25 MeV and a 10-day irradiation. This is in reasonable agreement with the initial estimate given in Section 2.4.1. It is not possible to use power as high as 10 to 20 kW on a single 3 mm thick converter target and a single one-gram radium target. The converter target is generally made from several pieces of tantalum with water cooling on both the front and back to provide high-efficiency cooling. The radium target, encapsulated with a protective material such as aluminum, can be produced in multiple segments, separated by sufficient space to water-cool each surface. Fluka was used to evaluate several possible set-ups of multiple radium targets with a converter of three one-mm pieces of tantalum. Figure 2 shows examples of two configurations that were evaluated. The left-hand side shows a configuration of three one-mm thick tantalum converter targets separated by 1.3 mm followed by four radium targets of ¼-gram each separated by 1 mm. The radium targets are 10-mm diameter by 0.63-mm thick based on using pure radium of density 5 g/cm$^3$. There is a one-mm beryllium or titanium window at both the front and back of the targets. The electron beam used was Gaussian with a full width at half maximum (FWHM) of 7 mm.

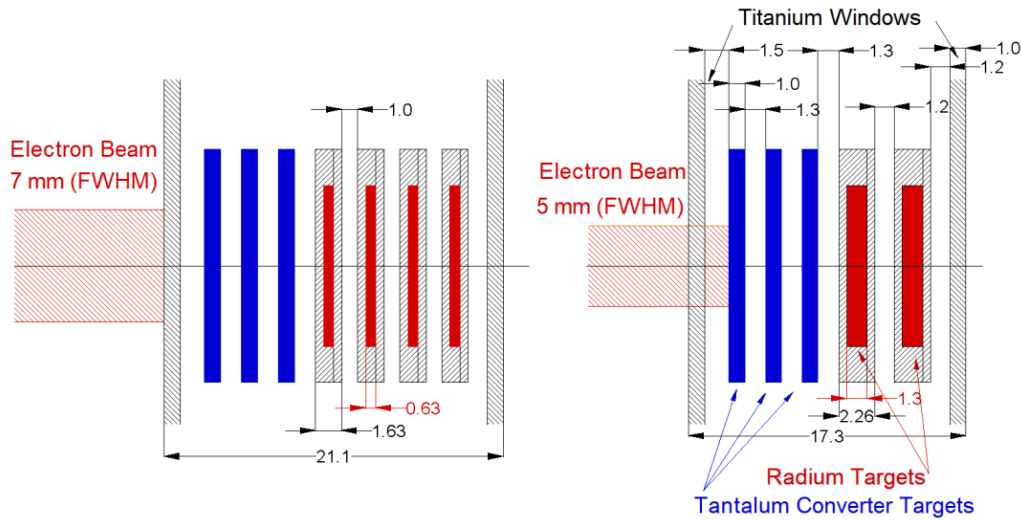

Figure 2. Target configurations evaluated with the Monte Carlo code Fluka.

The calculated yield of $^{225}$Ra was reduced to just over 50% of the optimized yield and several changes were made to improve upon that number. Figure 3 shows the calculated fluence of electrons at the entrance to the beam window for the 7-mm diameter electron beam and the fluence of photons of energy greater than 5 MeV at the exit of the third tantalum target and the exit of the fourth radium target for 25 MeV electron energy and 20 kW. The solid vertical line shows the radius of the radium target and the dashed line shows the width of the square tantalum sheets or the aluminum encapsulating the radium targets. The fluence of bremsstrahlung exiting the fourth radium target has been reduced substantially from the angular spread of the bremsstrahlung with distance and from the scattering and absorption in the other three targets.

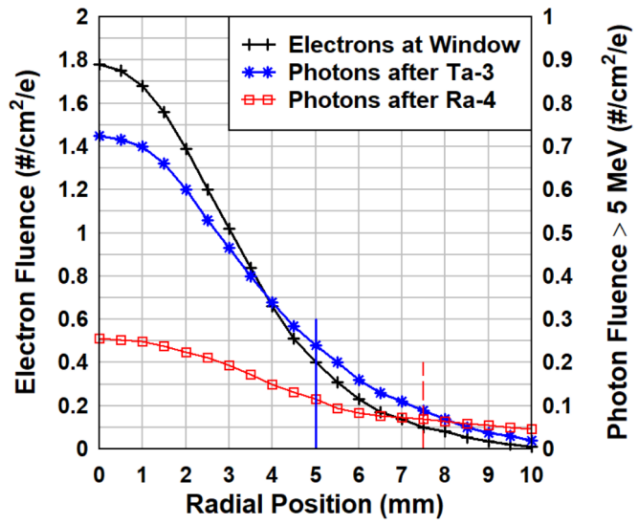

Figure 3. The electron and photon fluence at the entrance to the window (electron fluence) and at the exit of the third tantalum disk (photon fluence) and the fourth radium target (photon fluence) for the configuration shown in the left side of Figure 2.

The right-hand side of Figure 2 shows an example of two ½-g radium targets with the same windows and converter target as used in the four-target arrangement. The radium targets are 1.3 mm thick by 10 mm diameter and the spacing between the targets has been increased a small amount because of the higher power produced in the targets. The beam diameter was also reduced to 5 mm for this case. The changes were made to increase the fluence of photons on the radium targets.

Figure 4 shows the yields (EOB) of both configurations illustrated in Figure 2 for a 20-kW electron beam of 25 MeV and a 10-day irradiation. It also shows the case of using a 5-mm diameter beam with the configuration of four radium targets shown in Figure 2. 20 kW has been chosen because there was substantial experience at the Oak Ridge Electron Linear Accelerator (ORELA) facility of using an electron beam power of 50 kW with a beam width of 10 mm and water-cooled tantalum plates of varying thickness starting at 1.5 mm [17]. That target worked reliably for many years with both an aluminum and beryllium housing. The different configurations are shown because it may be necessary to limit the target power for low-thermal conductivity materials such as radium nitrate or radium sulphate. In that case the optimal yield of $^{225}$Ra may need to be produced by using three or four targets.

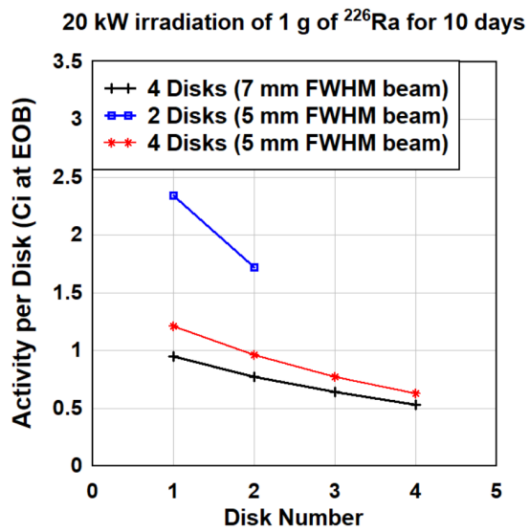

Figure 4. Yields of $^{225}$Ra from a 10-day irradiation at 20 kW on each of the two configurations shown in Figure 2. A third calculation with four radium targets (disks) and a beam diameter of 5 mm is also shown.

Table 2 shows the summary of the total calculated yields of $^{225}$Ra for each of the configurations in Figure 4 and the percentage yield compared to the single converter and single radium target, noted as "tight geometry".

Table 2. Summary of the total calculated yields of $^{225}$Ra for each of the configurations in Figure 4 for a 10-day irradiation at 25 MeV and 20 kW.

| Configuration | Beam Width (mm FWHM) | Activity Ci (EOB) | % of single target yield |
|---|---|---|---|
| Tight Geometry | 5 | 5.36 | 100 |
| Two Targets | 5 | 4.07 | 76 |
| Four Targets | 5 | 3.58 | 67 |
| Four Targets | 7 | 2.90 | 54 |

The Fluka calculations were used to provide estimates of the power dissipated in each of the various components that make up the target configuration. Table 3 shows the results of those calculations.

Table 3. Energy dissipation in each component of the target. Note that almost half of the incident beam power leaves the target area.

| Configuration | Beam Diameter (mm) | Element | MeV/e | Power[1] (kW) |
|---|---|---|---|---|
| Two Ra Targets | 5 | Window (Be) | 0.267 | 0.214 |
| | | Ta-1 | 2.64 | 2.12 |
| | | Ta-2 | 3.46 | 2.77 |
| | | Ta-3 | 3.06 | 2.45 |
| | | Ra-1 | 0.465 | 0.372 |
| | | Ra-2 | 0.237 | 0.190 |
| | | Al-1 | 0.496 | 0.397 |
| | | Al-2 | 0.305 | 0.244 |
| | | Water | 2.253 | 1.803 |
| Total | | | | 10.557 |
| Four Ra Targets | 7 | Ra-1 | 0.204 | 0.163 |
| | | Ra-2 | 0.123 | 0.098 |
| | | Ra-3 | 0.077 | 0.062 |
| | | Ra-4 | 0.049 | 0.039 |
| | | Al-1 | 0.425 | 0.324 |
| | | Al-2 | 0.281 | 0.225 |
| | | Al-3 | 0.195 | 0.156 |
| | | Al-4 | 0.130 | 0.104 |
| | | Water | 2.95 | 2.36 |
| Total | | | | 10.86 |
| Four Ra Targets | 5 | Window (Ti) | 0.670 | 0.536 |
| | | Ra-1 | 0.243 | 0.194 |
| | | Ra-2 | 0.143 | 0.115 |
| | | Ra-3 | 0.088 | 0.071 |
| | | Ra-4 | 0.055 | 0.044 |
| | | Al-1 | 0.442 | 0.337 |
| | | Al-2 | 0.296 | 0.237 |
| | | Al-3 | 0.202 | 0.162 |
| | | Al-4 | 0.133 | 0.107 |
| | | Water | 1.820 | 1.456 |
| Total | | | | 10.771 |

[1]Component power for a 20-kW electron beam power

For the case of four radium targets the maximum power in the radium is only 163 W with the 7-mm diameter beam and 194 W with a 5-mm diameter beam. The target is cooled on both sides by high-velocity water. The aluminum cladding on the first target absorbs 324 W but that is spread over an area that is 2.9 times greater than the radium target and the high thermal conductivity will reduce the temperature increase in the aluminum.

The $^{225}$Ra yield was calculated for the two-radium-target configuration shown in Figure 2, as a function of electron-beam energy. Figure 5 shows the results of the calculations. There is a significant increase between 20 and 25 MeV and modest increases for higher energies. If the irradiation time increases from 10 to 15 days, there is also an increase of 1.35. As an example, a 15-day irradiation at 30 MeV would produce 6.1 Ci (EOB) of $^{225}$Ra.

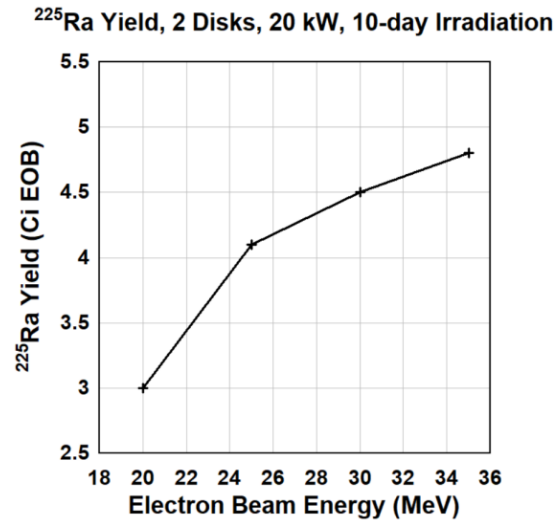

Figure 5. $^{225}$Ra yield (EOB) from two ½-gram radium targets at 20 kW beam power irradiated for 10 days, versus the electron beam energy.

### 2.3.3 $^{225}$Ac Yield

Section 2.3.2 shows the yield of $^{225}$Ra for several target configurations and irradiation times. $^{225}$Ac is building up throughout the irradiation in parallel with the $^{225}$Ra and continues to increase after completion of the irradiation. Figure 6 shows an example of a 10-day irradiation at 20 kW and 25 MeV on the two-target configuration of Figure 2. The $^{225}$Ac buildup is calculated using equation 5.44 from the CANTEACH [18] lectures. It reaches a broad maximum between about 15 to 30 days. The red square symbols show the case of milking about 2 Ci at 15 days (5 days post EOB) and waiting for another 15 days before a second milking of the target that produces another 1.3 Ci. For 100% extraction efficiency, the two extractions of $^{225}$Ac lead to a yield of about 80% of the yield of $^{225}$Ra at EOB. A third milking after another 15 days would yield another 0.65 Ci leading to an extraction of nearly as much $^{225}$Ac as the amount of $^{225}$Ra produced EOB.

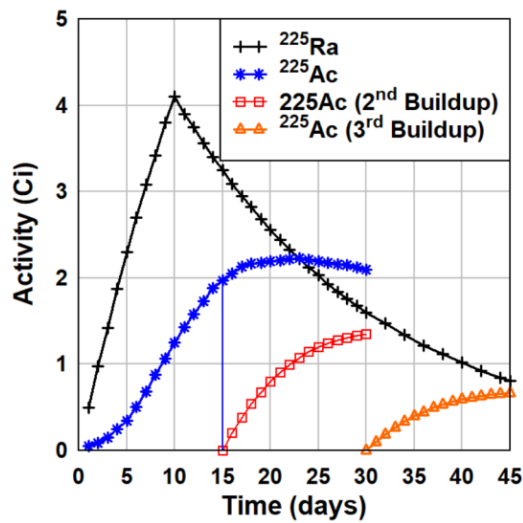

Figure 6. Build-up and extraction of $^{225}$Ac from an irradiated radium target. The vertical line shows extraction at 15 days after the beginning of the irradiation and the red squares shows the build-up of additional $^{225}$Ac. The orange triangles show the buildup of additional $^{225}$Ac during another 15-days of decay of the $^{225}$Ra.

The number of extractions of $^{225}$Ac from the irradiated radium target may depend on the degree of difficulty of the extraction from the target including potential losses of valuable target material.

### 2.3.4 Production of $^{224}$Ra

$^{224}$Ra will pose significant challenges during target processing post EOB because it will reach near equilibrium during a 10- to 15-day irradiation. One of its significant decay products is $^{208}$Tl which has a 3 minute half life and decays with the production of 2.6 MeV gamma rays. Fluka has been used to calculate the yield of $^{224}$Ra for the four-radium target configuration (see Figure 2) for the conditions of 20 kW electron-beam power, a 10 day irradiation and a beam profile of 5 mm (FWHM), as a function of electron beam energy. Figure 7 shows the results of those calculations and a comparison with the activity (EOB) of $^{225}$Ra.

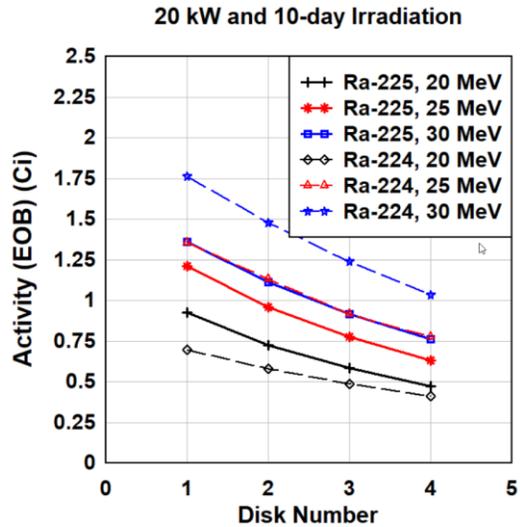

Figure 7. Comparison of the yields of $^{224}$Ra and $^{225}$Ra for four 250-mg radium targets and a 5 mm (FWHM) beam profile at three electron-beam energies. The yields are given as the activity per disk for a 10-day irradiation at 20 kW. The solid lines show the results for production of $^{225}$Ra at three different energies and the dashed lines for production of $^{224}$Ra for the same conditions.

The total yield of $^{224}$Ra is 2.2, 4.8 and 5.5 Ci at 20, 25 and 30 MeV, respectively.

**2.4 Design Example of a High-Power Radium Target and Target Chamber**

**2.4.1 Radium Target Design**

Section 2.1 outlines some of the advantages of radium targets used with an electron beam. The target material can be encapsulated in material that is much thicker than could be used with protons. This provides a high safety margin when using a target material such as radium that is highly radiotoxic and decays with the production of radon that is a gas at room temperature. The initial target material such as one gram of radium can be divided into smaller targets without substantial loss of yield. This reduces the activity of each of the smaller targets during target preparation and after irradiation and reduces the power that each target must dissipate during irradiation.

Figure 8 shows an example of the high-power radium target used in the two-target configuration of Figure 2. It uses two sections of high-purity aluminum metal to encapsulate the radium. Figure 8a shows a top view of the target. The aluminum encapsulation is 15 mm by 15 mm and the diameter of the radium is 10 mm. Figure 8b shows the details of the three sections. The base is made from 1.77 mm thick aluminum with the central 10-mm reduced in thickness to accept the radium target, the radium target is 10 mm diameter by 1.28 mm thick for metallic radium and the cover is made from a ½ mm thick sheet of aluminum. It is proposed to fuse the two aluminum sections using a cold weld under a rough vacuum so that the plates would collapse on to the target material when exposed to atmospheric pressure. High-pressure cooling water would increase the contact pressure and aid in heat transfer to the cooling water.

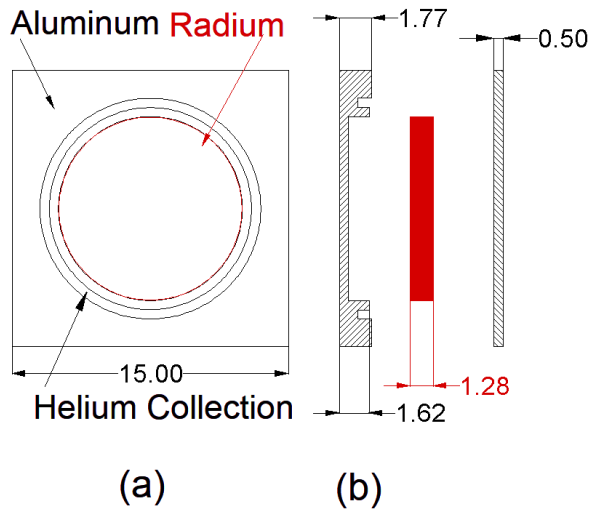

Figure 8. Example of a high-powered radium target for electron irradiation. Dimensions are in mm.

The aluminum cladding used for the base has a channel surrounding the radium that is used as a reservoir for helium produced by the large amount of alpha decay during irradiation and post-irradiation. Section 2.6 provides an estimate of the volume from this source. The aluminum encapsulation is sufficiently thick to withstand about 10 atmospheres of internal pressure and remain well within the yield strength of aluminum. This is substantially higher than the expected pressure increase. The top of the channel that surrounds the target region is about 0.15 mm lower than the cover plate to enable the helium to pass from the target to the channel. This also helps to separate the top from the base after irradiation and is discussed further in Section 2.9.

If the target is a salt of radium it may be preferable to evacuate the target and backfill with dry helium to provide extra thermal conduction from the target material to the aluminum cladding. The target material should be heated before it is encapsulated to reduce moisture. This will reduce the chances of buildup of radiolysis products inside the cladding during irradiation that may be corrosive and increase the internal pressure.

**2.4.2 Target Chamber Design**

Figure 9 shows an example of a target chamber that holds both the high-power converter target and two radium targets such as shown in Figure 8. This target chamber is modelled on the high-power neutron target used at the Oak Ridge Electron Linear Accelerator (ORELA) neutron source [17]. That target used about eight rectangular sheets of tantalum of increasing thickness from 1.5 mm to 7.6 mm with gaps between each sheet for water cooling. The tantalum sheets were contained in slots in either a beryllium or aluminum housing and were used as a 50-kW thick-target for the ORELA pulsed neutron source. The beam diameter was not measured with a beam scanner but was specified as about one cm (FWHM). This target worked reliably for many years with either housing. The high-power window was just the material of the housing and was water-cooled by the water used to cool the tantalum target plates.

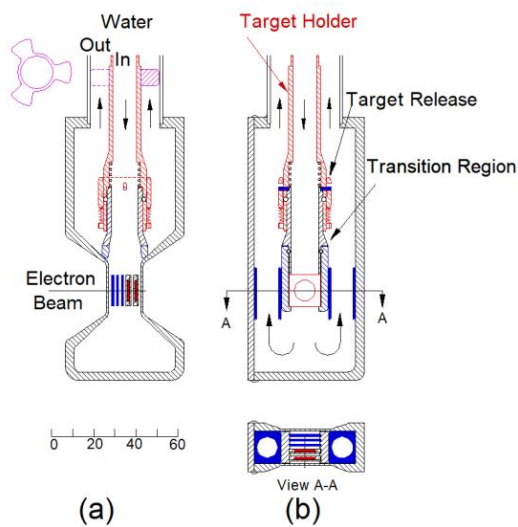

Figure 9. Example of a high-power target chamber that contains both an optimized converter target and an assembly of two radium targets for electron irradiation. Dimensions are in mm.

The target is cooled by high-velocity water in the center tube that passes through the spaces between the plates of the converter and isotope targets. The electron-beam window is machined as part of the target chamber and the chamber is shaped to stream the input cooling water through the two targets as shown in Figure 9a. The channels shown in Figure 9b are used to direct the cooling water back into the co-axial return channel for the cooling water. There is a three-lobed spacer at the top of Figure 9a that is used to center the target as it is inserted. The isotope targets are held in slots of a removable target assembly that can be separated into two sections once the target assembly has been moved to a hot cell.

In the region noted as "transition region" the geometry of the cooling line changes from circular in the input tube to rectangular throughout the remainder of the target assembly. The target assembly can be easily removed from the housing and can be released once it is clear of local shielding and deposited into a shielded flask during remote handling operations. The target assembly is held in the target holder by a spring-loaded ball-lock assembly. When the target assembly is withdrawn from the target chamber to a location above the iron shielding (see Section 2.5), the target release ring is actuated and the target assembly is released from the target holder. The target chamber material could be made from aluminum, beryllium or titanium. The chamber is closed by welding a plate on one side after the internal details are machined as shown in Figure 9b.

The target in Figure 9 uses three one-mm tantalum plates and two radium targets, all captured in a single aluminum holder to minimize the distance between the converter and isotope targets.

The close coupling of the converter targets and the isotope targets increases the yield of $^{225}$Ra for a given electron beam power and profile as shown in Section 2.3.2. The tantalum will be removed along with the radium targets during each irradiation. $^{181}$Ta is the only stable isotope of tantalum and the main photonuclear reactions are (γ,xn) to other tantalum isotopes. Table 4 shows the main decay characteristics of the three tantalum isotopes that will be produced. They all decay by electron capture (EC) that leads to the production of low-energy X-rays. Two ($^{180}$Ta and $^{178}$Ta) have short half lives and will be decayed substantially by the time the radium targets need to be removed from the target holder for processing. $^{179}$Ta will reach about 1% of saturation during a 10-day irradiation or about $2 \times 10^9$ Bq (55 mCi) for all three pieces and with the low-energy X-ray should not be challenging to handle post irradiation.

Table 4. Characteristics of isotopes produced by irradiating a tantalum converter target.

| **Reactions** | (γ,n) | (γ,2n) | (γ,3n) |
|---|---|---|---|
| Half-life | 8.15 h | 1.82 y | 9.3 min |
| Decay | EC (85%) and β$^-$ (15%) | EC (100%) | EC (100%) |
| γ-Energy (keV) | 54.6 (20%) | 54.6 (12.6%) | 54.6 (22%) |
|  | 55.8 (30%) | 55.8 (21.8%) | 55.8 (38%) |

There will also be some $^{182}$Ta produced by neutron capture from the background of neutrons produced by the photon irradiation of the tantalum and radium and moderated by the local shielding. This was modelled using the same approach as used in Section 2.5 to calculate the production of $^{227}$Ac produced by neutron capture on the radium target. The results were a yield from all three pieces of tantalum of $9.7 \times 10^8$ Bq (0.026 Ci) EOB. The half life is 114.7 d and it decays via multiple gamma rays with the most intense in the energy range from 1.1 to 1.2 MeV. The unshielded radiation field at one meter from this source calculated using MicroShield [10] is 18 mR/h and with 5 (7.5, 10) cm of lead it is 2.0 (0.5 and 0.1) mR/h. This should not significantly increase the handling challenges of removing the tantalum to a shielded container.

### 2.4.3 Target Handling

The irradiated targets of radium will have sufficient radioactivity that they will need to be shielded by about 15 cm of lead or the steel equivalent for significant occupation by radiation workers and about 10 cm of lead for limited operations. It would be useful if most of the target handling steps could be done remotely in a largely automated process. This section proposes several concepts to help enable such processes.

The target assembly shown in Figure 9b would be loaded remotely into a lead flask of about 10 cm thickness positioned below the target assembly that has been raised to the appropriate location and the ball-lock is activated to drop the target holder into the flask. The flask is moved to one side and a lid is installed. A second flask is moved into position and a new target is installed in the reverse procedure. Once the lid is installed the flask can be manually or remotely moved to a hot cell for further operations.

Figure 10 shows one potential approach to opening an irradiated target that could be easily automated. The irradiated target is held on a precision holder and moved in place under a shear. The shear penetrates the cover plate just inside the helium channel until a vacuum-actuated cup contacts the sheared lid and a rough vacuum is applied. The assembly is then lifted with the sheared lid attached. The small space between the cover plate and the base in the annular region between the target and the helium channel should prevent any cold welding at that location. By the time the target is to be opened there will be several atmospheres of helium that will help separate the cover sheet from the target material. The precision target holder would then be used to move the target to the next station for dissolution of the target material.

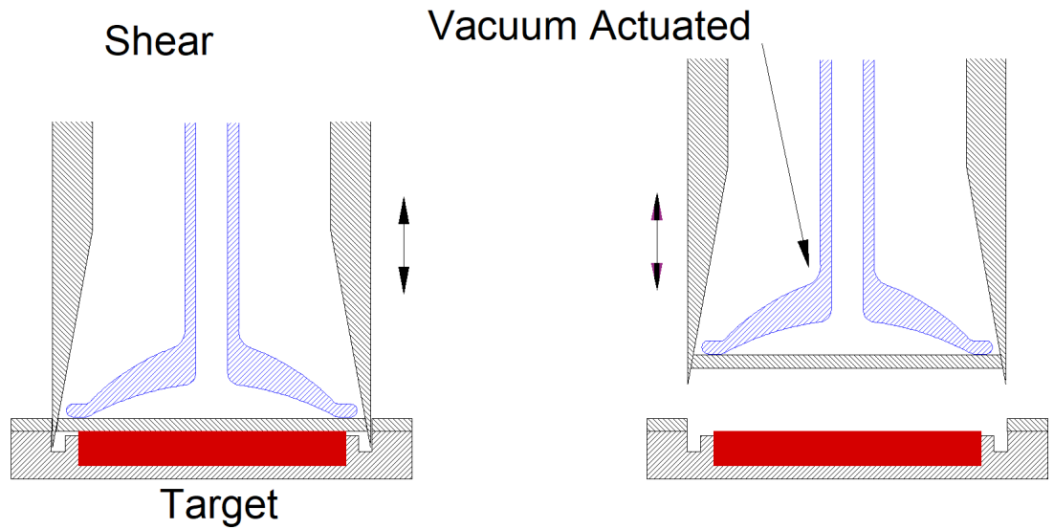

Figure 10. Opening an irradiated target.

The radium target shown in Figure 10 is a generic representation of the target. This basic target configuration can be used with many of the radium target designs that have already been developed for cyclotron targets. Examples include using multiple thin (15 μm) sheets of aluminum with radium electrodeposited on each sheet [19]. Reference 20 proposes evaporating a solution of a radium salt on a surface of an aluminum cup. This method can produce thicker targets than can be produced by electrodepositing a radium layer and could be readily adapted to the target base shown in Figure 8. It should be possible to produce a reasonably uniform electrodeposited layer of radium on a sheet of high-purity aluminum with a moderate temper that could then be punched into 1-cm diameter disks. To reduce the waste and contamination of the die, the aluminum could be masked on both sides with a pattern of about 9.7-mm diameter circles. This should leave a clean shear line for the punching process. It should be possible to shear all the targets in one operation to reduce the handling of the targets and reduce stress in the aluminum plate. Electrodepositions on both sides would lead to more target material per aluminum plate.

This target configuration could also be adopted to use fine powders of several radium salts. Figure 11 shows an example of using materials such as radium nitrate or radium carbonate as a fine powder on the same basic target configuration with the bottom modified to provide small aluminum channels that would provide additional cooling to the target material. The fine channels could be reproducibly produced by techniques such as "plunge" electric discharge machining (EDM) on the target base.

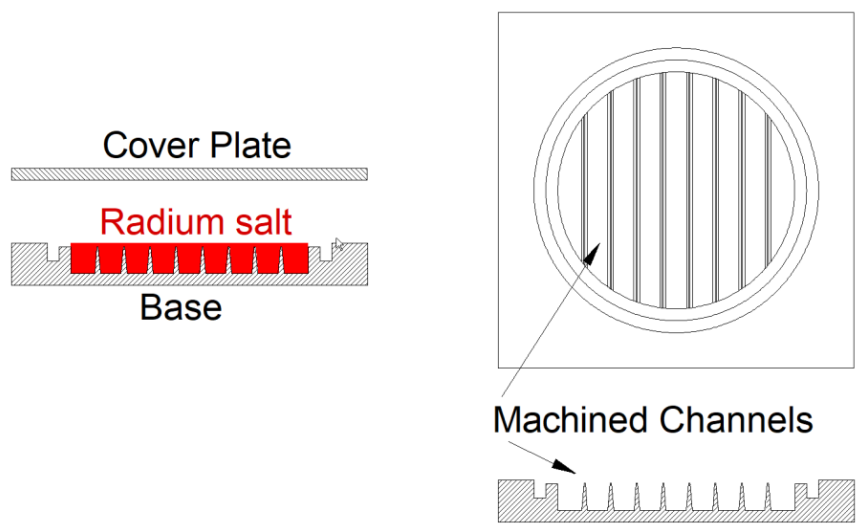

Figure 11. Radium target designed for powered radium salts.

The challenges of controlling the use of fine powders of a radioactive material to prevent spills and equipment contamination could be largely controlled by an automated process that contains the material in a small volume system with double containment until the cover plate is installed. The inner surfaces of the containment system should be capable of being rinsed as required, by a solvent such as nitric acid that can dissolve the radium salt in use. The liquid would be processed to recover the radium.

It should also be possible to use some of the aluminum dispersion target technology that has been developed for uranium targets to improve the thermal conductivity of the target material.

## 2.5 Reduction of $^{227}$Ac Production

At an electron beam power of 20 kW, the converter target will produce about 5 x 10$^{12}$ neutrons per second in addition to the neutrons produced from the radium targets [11]. These have an energy spectrum like the fission neutron spectrum with a peak around one MeV. $^{226}$Ra has a high thermal neutron capture cross section of 14 b [21] leading to $^{227}$Ra that β-decays quickly (half-life of 42 min) to $^{227}$Ac. $^{227}$Ac has a half-life of 21.8 years and there will likely be requirements for a low ratio of $^{227}$Ac:$^{225}$Ac [22]. The local thermal neutron background could be reduced considerably if no local shielding were used around the target chamber and the neutrons produced by the targets and the intense bremsstrahlung irradiated the target room walls. However, intense bremsstrahlung produces a substantial amount of ozone in the air and radioactive air products [11] and the neutrons produce high residual activity in the target room that would limit access. This can be greatly reduced by using local high-density shielding around the target, but the local shielding will also moderate the neutrons and might produce sufficient thermal neutrons to produce levels of $^{227}$Ac that are unacceptable.

Kukleva et al. reported a yield of $^{227}$Ac of 0.1 Ci/g of $^{226}$Ra in a thermal neutron flux of 1.1 x $10^{14}$ n/s/cm$^2$ for 11.7 days [23]. This is about the same irradiation period as would be used to irradiate one gram of $^{226}$Ra with electrons producing the order of 2 Ci of $^{225}$Ac after the first milking from the radium target. Note that a second milking of the target as illustrated in Figure 6 should have much lower concentrations of $^{227}$Ac because it would be removed during the first milking. To reduce the yield of $^{227}$Ac to a value the order of about 1 x $10^{-6}$ of the yield of $^{225}$Ac, would require reducing the thermal neutron flux at the radium target to a value lower than about 2 x $10^9$ n/s/cm$^2$. The production of $^{227}$Ac from an irradiation of radium in a typical research reactor would include contributions from the resonance integral energy region that should be about the same as produced by the moderated photo-neutron spectrum.

Figure 12 shows a target chamber assembly with local shielding that reduces the thermal and resonance integral neutrons at the region of the target assembly by a large amount. This figure shows a side view of the assembly. The vacuum chamber is a cylindrical vessel of 30 cm diameter that connects to the linac vacuum system. The top and bottom of the vacuum chamber are flat plates welded to the cylinder. The target chamber containing the converter and isotope targets shown in Figure 8 is centered in the vacuum chamber. It is sealed at the top of the vacuum chamber and can be removed from the vacuum chamber for replacement if a window fails. The vacuum chamber is surrounded by a cylindrical vessel of 15 cm thickness that is cooled by water that could contain a strong neutron poison. High-energy neutrons from the target region pass through the vacuum chamber without interaction and enter the tank with water loaded with the neutron absorber. The water moderates the neutrons and the poison, with a high thermal neutron cross section, captures thermal and some resonant-integral neutrons.

The intense bremsstrahlung, especially in the forward cone, passes through the water shield and interacts with the steel shielding that attenuates the bremsstrahlung by about three decades from 90$^o$ with respect to the beam direction back to the entrance of the beam into the shielded enclosure. In the forward cone the reduction is at least four decades. The 15 cm of water is about ½-$X_o$ thick and will absorb significant energy from the bremsstrahlung with only a small contribution of neutrons because the cross section for neutron production is low. Figure 12 also shows a cooled aluminum block in the forward direction. It will produce less than 40% of the neutrons than would be produced in iron shielding in that region and can be easily cooled to remove the heat produced by the intense photon beam. Some of the neutrons produced in the steel will be captured by the water plus neutron poison in the central region while others are captured in the iron with a capture cross section of 2.7 barns [24]. The (γ,n) products in iron produce either short half-life products such as $^{56}$Mn with a half life of 2.6 h or products such as

[55]Mn that decay by electron capture with little penetrating radiation [24]. The products of neutron capture of iron isotopes are mostly other stable iron isotopes. This reduces the residual radiation in the isotope production room to manageable levels soon after the electron beam is turned off. The remote handling station described in Section 2.4.3 would be located just outside the steel shielding to remove and replace the radium targets.

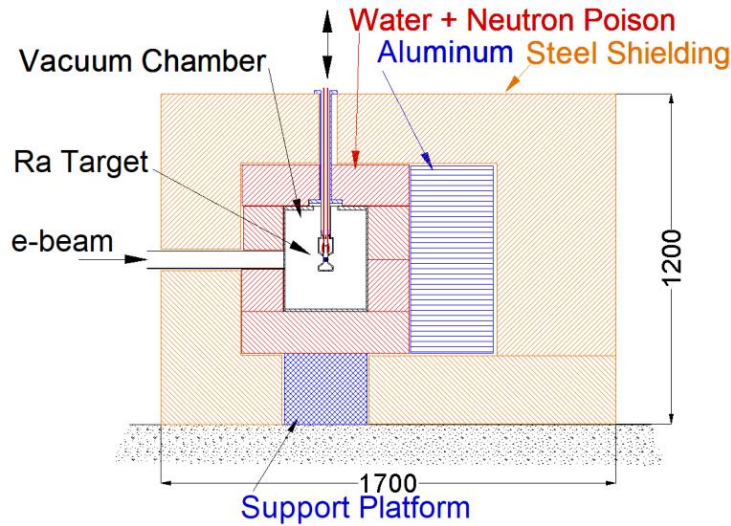

Figure 12. Schematic of a target assembly that reduces the production of [227]Ac to acceptable levels. Dimensions are in mm.

There are several soluble poisons used in the nuclear power industry that have demonstrated the capability to function in the harsh environment of a nuclear reactor. Boron, in the form of boric acid, is used as a neutron poison. [10]B, with an isotopic abundance of 20%, has a high thermal neutron cross section. The [10]B(n,α)[7]Li reaction has a thermal neutron cross section of 3835 barns [21]. Boron will also attenuate neutrons in the resonance integral range because it maintains a high absorption cross section throughout the energy range to about one MeV [21]. Boric acid is a weak acid with a maximum concentration of 4.7 g per 100 g of solution of boric acid at room temperature [25] and the chemical makeup is $B(OH)_3$ (molecular weight of 61.8 g/mole). The combination of 20% of [10]B and the maximum of 47 g of boric acid or 47 x 10/61.8 = 7.6 g of natural boron per 1000 g limits the amount of [10]B in solution to about 1.5 g of [10]B per 1000 g of solution.

Fluka has been used to calculate the neutron spectrum in the region of the radium targets. Figure 13 shows the neutron flux for 20 kW of 25 MeV electrons on the two-radium target configuration. This figure shows the flux for the case with just water in the cylinder surrounding the target and with the addition of 4.7% boric acid. There is a significant reduction in the neutron flux throughout the low-energy region. At thermal energy of 0.025 eV, the flux has been reduced from about 1.5 x 10$^9$ n/s/cm$^2$ to 1.5 x 10$^8$ n/s/cm$^2$. This is below the level to reduce the yield of [227]Ac to less than 1 x 10$^{-6}$ of the yield of [225]Ac if the yield were only from thermal neutrons.

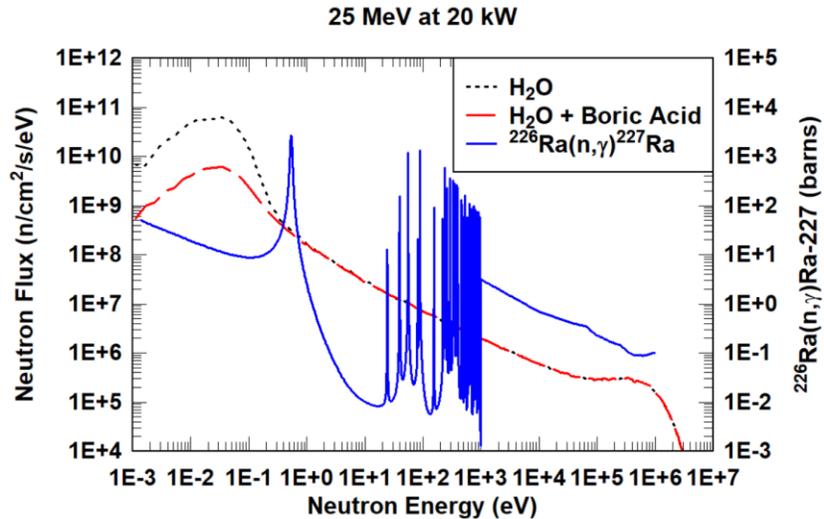

Figure 13. Neutron flux at the radium target for the experimental setup in Figure 10 with and without boron in the water annulus. The cross section for $^{226}Ra(n,\gamma)^{227}Ra$ [26] is also shown referenced to the right-hand y-axis.

Figure 13 also shows the cross section of $^{226}Ra(n,\gamma)^{227}Ra$ [26]. There is a significant cross section (100's of mb) for neutron energies from below 1 keV to beyond 1 MeV and there is a substantial high-energy flux of neutrons near the target that needs to be considered. The neutron capture cross section was integrated with the neutron fluence to get an estimate of the effect of the high-energy neutrons. Figure 14 shows the results of those calculations for the case with water (no poison) in the annular tank. The red solid line is the integrand as a function of neutron energy. The black dashed line is the value of the integral up to a given energy. This gives the reaction rate for a single $^{226}Ra$ atom. Up to 1 eV, the integral is 1.08 x $10^{-13}$ /s, while up to 20 MeV it is 1.91 x $10^{-13}$ /s. High energy neutrons contribute about 40 % to the yield. One gram of $^{226}Ra$ contains 2.67 x $10^{21}$ atoms so the reaction rate for production of $^{227}Ra$ from the entire energy spectrum is 5.1 x $10^{8}$ atoms/s. $^{227}Ra$ rapidly β-decays (half-life of 42 min) to $^{227}Ac$. The activity of $^{227}Ac$ at the end of a 10-day irradiation is:

5.1 x $10^8$ x (1-exp(-λt)) = 4.4 x $10^5$ Bq, modestly higher than the estimate based on the thermal neutron flux alone. This value is about 6 x $10^{-6}$ of the calculated yield of 2 Ci (74 GBq) of $^{225}Ra$ or 6 parts per million.

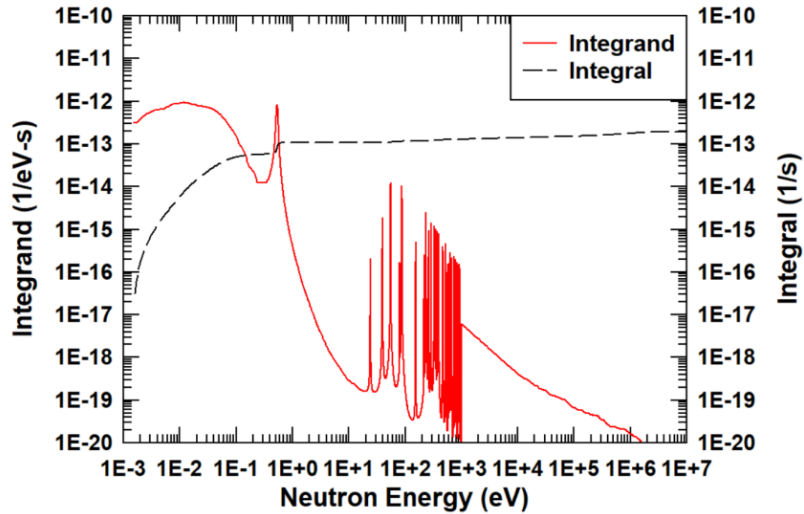

Figure 14. Product of the neutron capture cross section times the neutron fluence produced by Fluka. The red solid line is the integrand as a function of neutron energy. The black dashed line is the value of the integral up to a given energy.

To obtain an estimate of the minimum yield of $^{227}$Ac that might be obtained, calculations were done with Fluka with just the target chamber. Figure 15 shows the neutron spectrum at the target for the target chamber with either light-water or heavy-water cooling. The small quantity of light water in the target chamber is sufficient to produce a significant thermal neutron flux. The addition of a strong neutron poison in the shielding tank has no effect on this source of thermal neutrons. Heavy water has a much lower (about 12%) Macroscopic Slowing Down Power (MSDP) compared to light water and the small volume of cooling water does not moderate a significant number of neutrons to thermal energy with heavy water in the target chamber.

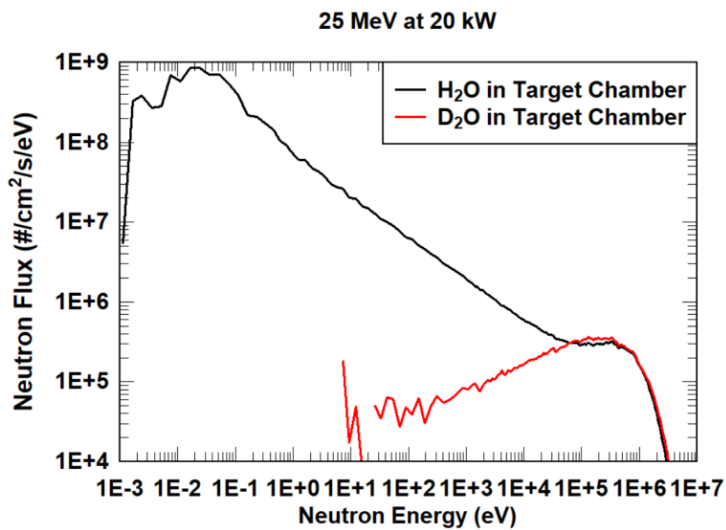

Figure 15. Neutron spectrum for the target chamber with no local shielding. The solid black line shows the case with light water in the chamber and the solid red line shows the case for heavy water.

Figure 16 shows the same calculation as shown in Figure 14 for the case of the target chamber cooled with heavy water. This provides an estimate of the lower limit of yield of $^{227}$Ac. The black dashed line is the value of the integral (up to 20 MeV it is $5.5 \times 10^{-14}$ /s) leading to a yield of $^{227}$Ac of $1.5 \times 10^8$ atoms per second or $1.3 \times 10^5$ Bq at EOB for 20 kW and 25 MeV and a 10-day irradiation. This is about 30% of the full shielding configuration with no neutron poison.

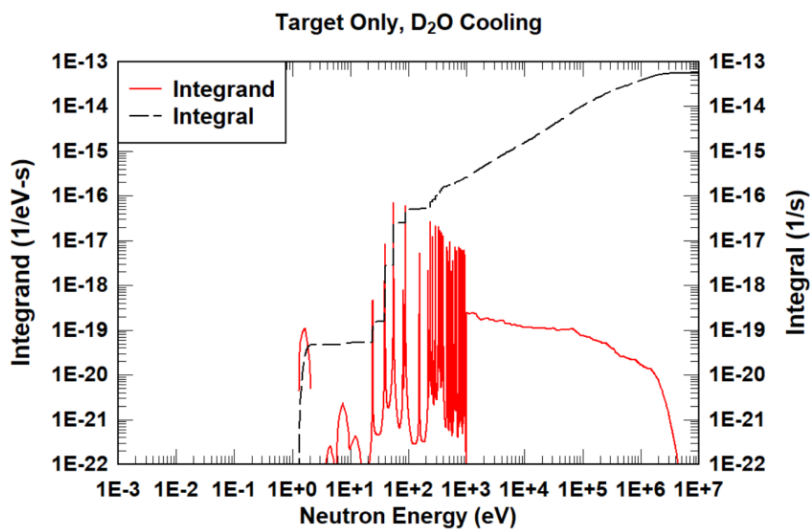

Figure 16. Product of the neutron capture cross section times the neutron fluence produced by Fluka for the beam dump cooled by heavy water. The red solid line is the integrand as a function of neutron energy. The black dashed line is the value of the integral up to a given energy.

This same methodology was used to calculate the production of $^{182}$Ta from neutron irradiation of the tantalum targets. The results given in Section 2.4.2 were for the yield from all three pieces of tantalum of $9.7 \times 10^8$ Bq (0.026 Ci) EOB for the case of light-water cooling and water in the annulus with no poison.

## 2.6 Helium Production during High-Yield Irradiations

The production of high yields of $^{225}$Ac is accompanied by co-production of large quantities of alpha particles that become helium atoms. The IAEA document, IAEA-Tecdoc-886 [27], contains the following "The decay of each atom of $^{226}$Ra yields five helium atoms formed from the alpha particles emitted in the decay chain. This generates overpressure in the sealed radium source (about 0.2 atmospheres per year for one gram of radium assuming a free volume of 1 cm$^3$) which facilitates leaking and spread of contamination". This is equivalent to 200 mm$^3$ at standard temperature and pressure (STP). For a 15-day irradiation and a 15-day decay period, this source alone would provide 16 mm$^3$ at STP or 8 mm$^3$ in each of the two radium targets.

An irradiation of a one-gram radium target for 10 days, at a reaction rate of about $2.1 \times 10^{11}$ /s for $^{224}$Ra (see Section 2.3.4) and $4 \times 10^{11}$ /s for $^{225}$Ra (see Section 2.3.2) will produce the order of 2 and $3.5 \times 10^{17}$ reactions, respectively, by 10 days of irradiation. $^{224}$Ra decay will reach near equilibrium by that time producing four alphas per initial $^{224}$Ra atom. $^{225}$Ra will β-decay to $^{225}$Ac that continues to decay with the production of four alphas. Towards the end of the irradiation there will be roughly three alphas produced per $^{224}$Ra atom and less than one alpha per $^{225}$Ra atom produced. This will produce the order of 1 x 10$^{18}$ atoms of helium in the two radium targets followed by another 4 x 10$^{17}$ atoms of helium during the decay for 15 days. It is difficult to predict the outcome of this high yield of helium nuclei. Some will remain trapped in the grain boundaries and other defects of the crystal structure and others will become free helium. There are many observations of damaged radium needle sources that suggest that a substantial fraction of the helium is released as a gas [27].

The total amount of helium produced during the proposed irradiation of 20 kW for 10 days is large if it is available as a gas.

One mm$^3$ of helium at STP = 2.7 x 10$^{16}$ atoms

Therefore, the amount of helium produced during irradiation could be the order of 38 mm$^3$ at STP if it were all available as free helium. Another 20 to 30% could be produced during the decay period depending on the milking scheme. With two targets this is reduced to about 25 mm$^3$ at STP in each target. The channels shown in Figure 7 for helium collection are about 15 mm$^3$ and are evacuated before irradiation. If the target material is metallic radium sealed under vacuum these channels should fill to about one atmosphere above standard pressure at room temperature and if substantially higher temperatures are produced throughout the target then the pressure will rise because there is only a fixed volume. The small pressure increase will be countered by the external water pressure of four to five atmospheres. If the target material is a salt of radium that is sealed with at least one atmosphere of helium for better heat transfer, the pressure rise will be to about three atmospheres at room temperature. A simple calculation of the stress on a fixed disk of ½-mm thick aluminum, supported at 12 mm diameter, shows that 10 atmospheres is still well within the yield of aluminum.

There will also be $^{222}$Rn produced by the decay of $^{226}$Ra in each of the irradiated targets. One curie of radium used as target material will produce one curie of radon in equilibrium with the radium. There will also be significantly larger quantities of $^{220}$Rn (half-life of 56 s) in equilibrium with the $^{224}$Ra produced during the irradiation. These will need to be captured in appropriate filtration in a closed hot cell used for all stages of target handling.

### 2.7 Residual Heat in an Irradiated Target

The decay heat in an irradiated radium target is much smaller than produced in fission targets used to produce $^{99}$Mo but needs to be considered during post irradiation target management, especially since there may be a high concentration of gaseous helium sealed in the target. The largest source of decay heat will be from the decay of nearly 5 Ci of $^{224}$Ra at EOB. It is nearly in secular equilibrium at that time and each $^{224}$Ra nucleus produces four alpha-decays, multiple beta-decays and decay energy of the order of 27 MeV [24].

27 MeV x 5 x 3.7 x 10$^{10}$ dps = 5 x 10$^{12}$ MeV/s = 0.8 W in two targets

$^{225}$Ra decays via an initial beta-decay of 0.1 MeV average energy. At EOB, there will be about 1.3 Ci of $^{225}$Ac and the other three alphas produced in the decay of $^{225}$Ac have much shorter half-lives than the initial decay and will be in secular equilibrium. There will be about 27 MeV in energy from this decay chain at EOB producing about 27 MeV x 1.3 x 3.7 x $10^{10}$ dps or 1.3 x $10^{12}$ MeV/s = 0.21 W.

The decay of the $^{226}$Ra target will be in secular equilibrium producing about 30 MeV at a decay rate of 3.7 x $10^{10}$ dps producing 1.1 x $10^{12}$ MeV/s = 0.18 W.

This will produce about 0.6 watt of decay energy in each of the two targets. The total target mass for each target is about 1.2 gram of aluminum plus the 500 mg of radium, or about 1.7 g. Aluminum has a specific heat of 0.9 J/g-$^{o}$C and radium has a specific heat of only 0.12 J/g-$^{o}$C so a 0.6 watt decay energy would produce about 0.6 $^{o}$C/s temperature rise in an isolated target. It will be necessary to provide some cooling during the entire decay period after EOB to avoid a substantial temperature rise accompanied by an increase in the helium gas pressure. This could be an active source such as an air stream or a small amount of water or a passive cooling such as heat sinking to a larger thermal mass.

## 2.8 Conceptual Layout of an Electron Accelerator Isotope Facility

There are several commercial vendors that produce electron accelerators that are specifically designed to produce medical isotopes. Both Mevex [28] and IBA [29] produce commercial electron accelerators for isotope production based on robust technology used at lower electron energies for multiple industrial uses such as sterilization. These accelerators are designed for long periods of operation with little maintenance required. Niowave [7] also produces a superconducting electron accelerator that it uses for production of $^{99}$Mo and $^{225}$Ac.

The peak photonuclear cross section is just under 50% of the cross section for proton production but higher-powered electron beams can compensate for some of that difference. The electron accelerators designed for isotope production can produce high-power electron beams, in the range of 50 to 400 kW, more than can be practically used.

Figure 17 shows a conceptual design of an electron accelerator facility to produce $^{225}$Ac, based on using a room-temperature linac. The linac is about three meters long by less than ½ meter wide and can be set on a stand that is between ½ and one meter above the floor. The target assembly (see Figure 10) can be used at zero degrees, in the same direction as the linac, to reduce the shielded volume by a significant fraction. However, there is a small risk that the beam window on the target chamber (see Figure 9) will fail during operation producing small metal particles that could travel into the linac interior, potentially damaging a cavity or the electron gun. There is a fast vacuum valve next to the linac to reduce this risk. The strong hermetic encapsulation on the radium targets should prevent the simultaneous failure of a radium target that would enable radioactive contamination of the accelerator and beam line.

The configuration shown in Figure 17 uses a bending magnet (either 90° or 270°) to direct the electron beam at 90° to the direction of the linac.  This reduces the risk of damage to the linac from a window failure by a large amount. This figure shows basic beam handling components to direct (steerers) and shape (quadrupole) the beam to provide a well-centered beam of a known profile at the isotope target.  The beam slits at the entrance of the target assembly are four tapered collimators that are designed to intercept some beam power during the beam setup.  A high-power beam dump is also shown on the zero-degree line.  This would be used for initial tune up of the linac before directing the beam to the isotope target.  This dump would be shielded with typically about 20 cm of lead to reduce the intense bremsstrahlung field directed at the facility wall.

There is an interior shielding wall (1/2-m thick) that separates the high-energy end of the linac and the bending magnet from the workspace near the target assembly.  There will be from about three to five percent beam loss near the end of the linac and in the bending magnet that will produce strong residual radiation fields from the copper after the accelerator has been turned off.  The shielding wall would be about two meters tall and not reach the ceiling of the shielded enclosure (about three meters high).  This enables the use of a light-duty, likely manually operated, crane that covers the interior space.  This would be useful when removing the shielding blocks around the target assembly and for other maintenance work.  The open space would be used for remote handling components and as a set-down area for the steel shielding pieces around the target assembly.

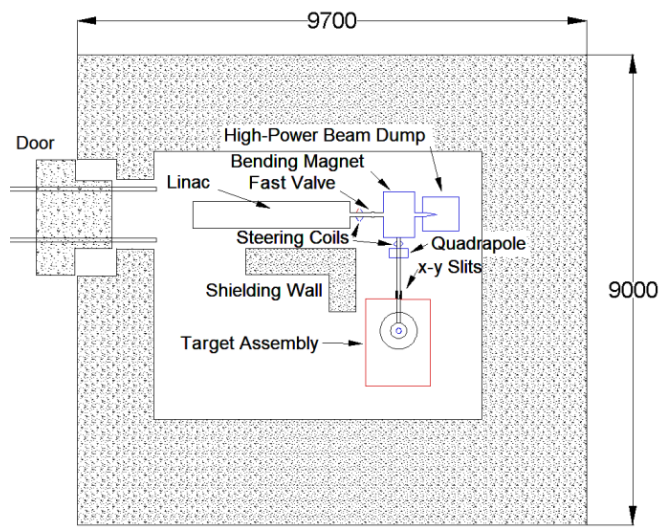

Figure 17.  Conceptual design of an electron accelerator isotope facility based on using a room-temperature linac as the accelerator. Dimensions are in mm.

There would be conventional construction outside the shielded walls that contain the high-power modulators and klystrons and a control area.  Hot cells for target handling could also be used outside of one of the shielded walls.  It would also be possible to transfer the target assembly (see Figure 9) remotely or manually in a shielded flask for transfer to another location for processing.

**2.9 Target Processing**

There was considerable work done on extracting and processing radium for many years during the first half of the 20$^{th}$ century. Radium was used in cancer treatment and other medical uses and to produce luminous paints for watches and other instruments. Reference 30 is a good summary of some of some of this work including references to previous publications. Reference 30 provides a procedure to separate actinium from radium. There are also patents and journal papers on separating $^{225}$Ac from $^{225}$Ra, for example, references 20 and 31. Reference 32 shows the separation of $^{225}$Ac from a more complex array of actinides and fission products. Maslov et al. [8] separated the $^{225}$Ac from $^{225}$Ra during the experiments using an electron beam to irradiate a radium sulphate or radium nitrate target. The preferred route will likely depend upon the final choice of the target material (see Section 2.4.3). Target choices that enable separation into multiple thin targets such as electrodeposited aluminum sheets would allow processing of small amounts of material and ease the handling challenges.

### 3. DISCUSSION

An electron-accelerator-based facility can be used to produce about four curies of $^{225}$Ra during a 10-day irradiation at 20 kW at 25 MeV. The $^{225}$Ra decays to $^{225}$Ac that can be milked from the radium targets 5 to 10 days after EOB, a second time about 15 days later and possibly another milking after an additional 15 days. High-powered electron accelerators are available commercially at competitive prices and the basic shielded facility design is quite straightforward. Bremsstrahlung penetrates windows and multiple targets without large power losses. This provides great flexibility in the target designs.

This paper proposes using two targets, each of 500 mg of radium target material, that are encapsulated in thick aluminum sheets to provide hermetic sealing and protection against overpressure by helium buildup during irradiation and decay. The target and target chamber design proposed should provide a yield of $^{225}$Ra that is about 75% of the yield of irradiating a single one-gram target of radium. The maximum power produced in the aluminum cladding and target material is about 600 W, cooled on both sides, with about ½ of the power produced in the aluminum that is easily removed. The target concepts can be initially tested with barium analogs to study possible thermal effects on the target material with much lower risk than using radium. It is also possible to use up to four 250 mg targets with a small reduction in yield (compared to two radium targets) because of the expanding bremsstrahlung beam. The multiple targets reduce the power per target, helium buildup and the activity per target during all phases of target handling. This may be the preferred choice, depending upon the target material.

The rugged target encapsulation also enables target production and post-irradiation processing at one location and irradiation at another. The requirement for 5- to 10-day decay after irradiation provides time for the co-produced $^{224}$Ra to decay by a factor of two to five before shipping to a processor, greatly reducing the external gamma field.

There are no competing reactions leading to other actinium isotopes except for the possibility of neutron capture in the radium target that leads to $^{227}$Ra that rapidly decays to $^{227}$Ac. A design is proposed that reduces the thermal neutron flux in the vicinity of the target to ensure that the ratio of $^{227}$Ac:$^{225}$Ac is as low as a few parts per million. The only co-produced reactions other than (γ,n), (γ,2n) and (γ,3n) are (γ,p) at less than 0.03 mb for 20 MeV photons [12] and (γ,f) at about 0.1 mb for 20 MeV photons. The (γ,p) reaction leads to $^{225}$Fr that rapidly decays to $^{225}$Ra. The (γ,3n) reaction increases to as high as 32 mb for photons of 22 to 24 MeV leading to $^{223}$Ra. This decays by an alpha decay and can be separated from the $^{225}$Ac in the same manner as $^{224}$Ra is separated.

The design concepts shown in this paper should require less development than may be required for other methods of high-yield production of $^{225}$Ac. The design of the converter target at 20 kW is quite conservative and with operations experience it may be possible to use higher powers, perhaps as high as 30 kW. It is based on the ORELA design [17] that was in use at much higher powers for many years. The radium target is also a conservative design. It should be capable of the few hundred watts on each individual target produced by a 20-kW electron beam. If there are challenges in using the two targets with the proposed 20 kW electron-beam then the target can be subdivided into more sections with only a modest loss in yield. The cladding is designed for extended operation in intense electron beams.

These results depend upon the cross section used in Fluka. Fluka does not use cross sections from a data base but instead it uses measured data to define the total photoabsorption cross section and then "event generators" are used to describe the de-excitation process. The only way to extract the cross section is to do a Monte Carlo "measurement". The results from this are shown in Figure 18. This shows the (γ,n) and (γ,2n) cross sections from the TENDL data base [12] and as calculated by the Monte Carlo measurement using Fluka. There is reasonable agreement between the two approaches for the (γ,n) cross section but the TENDL (γ,2n) cross section is significantly higher than the Fluka values. A higher value of the (γ,2n) cross section would lead to higher yields of $^{224}$Ra than predicted.

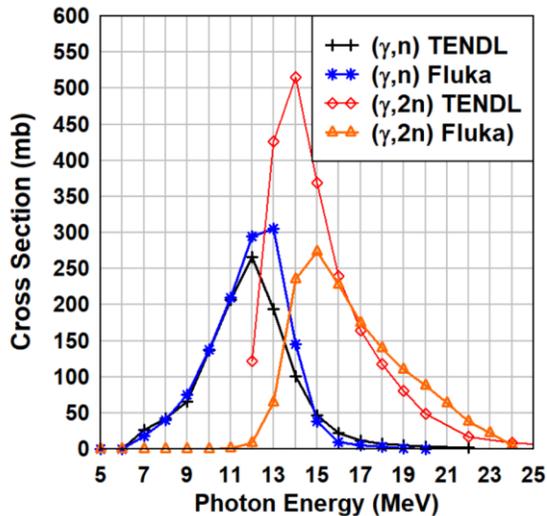

Figure 18.  Comparison of the TENDL and Fluka cross sections for (γ,n) and (γ,2n) for $^{226}$Ra.

In either case, the cross section for $^{226}$Ra(γ,n)$^{225}$Ra is a model calculation and may have a significant error.  However, the work by Maslov et al. [8] provides reasonable support for the cross-section values used in the Fluka Monte Carlo code.  They measured the production of $^{225}$Ac by irradiating two different 0.65 µg radium targets, one made from a mixture of powdered barium sulphate and radium sulphate and the other from radium nitrate, with bremsstrahlung produced by a 24 MeV electron accelerator at 15 µA [8].

They reported a yield of:

550 Bq of $^{225}$Ac/µA-h-mg $^{226}$Ra for both samples

The $^{225}$Ac was extracted from the $^{225}$Ra after a decay period of 18 days to maximize the yield of $^{225}$Ac at 44% of maximum activity of the parent $^{225}$Ra.

They compared that yield with the yield from a thin (7.16 mg of lead) and a thick (2.008 g of lead) lead chloride target.  The measured yields from those two experiments were:

34 Bq/µA-h-mg Pb for the thin target and

28 Bq/µA-h-mg Pb for the thick target

The lead results show good scaling from the small mg targets to much larger targets of several grams.  The lead yield was obtained from the gamma-ray decay of $^{203}$Pb produced by the (γ,n) reaction on the 1.4% abundant $^{204}$Pb isotope.  Most of the other more abundant isotopes of lead produce another stable lead isotope from the (γ,xn) reactions.

Fluka was used to calculate the yields of both the radium and lead experiments in the Maslov et al. paper [8].  The possible systematic errors that could make it difficult to obtain an absolute yield measurement from that experiment should be present in both the lead and radium measurements making it possible to obtain a better estimate by comparing the two measurements.  The (γ,n) cross section has not been measured for $^{204}$Pb but has been measured [16] for $^{206}$Pb, $^{207}$Pb, $^{208}$Pb and $^{209}$Pb and the peak value is between 500 and 600 mb.  Appendix 1 gives the details of the experimental setup and the results of multiple calculations using different electron-beam profiles. The results of the calculations are:

$^{225}$Ac – experimental results of 550 Bq of $^{225}$Ac/μA-h-mg $^{226}$Ra

- calculated results of 415 to 456 Bq of $^{225}$Ac/μA-h-mg $^{226}$Ra

about 75% to 83% of the measured value.

The variation depends upon the beam profile used in the calculation as shown in Appendix 1.

$^{203}$Pb – experimental results of 28 Bq of $^{203}$Pb/μA-h-mg of natural lead for the thick target

- calculated results of 22.1 to 23.8 Bq of $^{203}$Pb/μA-h-mg of natural lead for the thick target

about 79% to 85% of the measured value.

A comparison of the TENDL [12] values for (γ,n) for the three major fractions of lead ($^{206}$Pb at 24%, $^{207}$Pb at 22% and $^{208}$Pb at 52%) to the measured values [16] show that the TENDL values are about 80% of measured values, consistent with these observations.

Although there are no reported measurements of the $^{226}$Ra(γ,n)$^{225}$Ra cross section the reasonable agreement between the Fluka calculations and the measured data in the Maslov et al. paper for both lead and radium provide some assurance that the calculated yields in this paper are reasonable.

### 4. SUMMARY

There has been growing clinical evidence of the value of targeted alpha therapy for treatment of several cancers. The work has been slowed by the lack of availability of the key alpha emitting isotopes, especially $^{225}$Ac. There is an ongoing effort to produce new sources of $^{225}$Ac from several different accelerator-based routes. This paper proposes that a facility based on an electron accelerator of about 20 kW and an electron beam energy of 25 MeV could be a preferred route to producing high yields of $^{225}$Ra that decays to $^{225}$Ac. The reaction produces few contaminants other than $^{224}$Ra and $^{223}$Ra that can be separated from the $^{225}$Ac. The design concepts for a high-power converter target and segmented radium targets should lead to reduced development efforts compared to other approaches. The shielded target chamber has been designed to reduce contamination from $^{227}$Ac to a few parts per million of the yield from $^{225}$Ac. Although there are no reported measurements of the $^{226}$Ra(γ,n)$^{225}$Ra cross section the good agreement between the Fluka calculations and the measured data by Maslov et al. [8] for both lead and radium provide some assurance that the calculated yields are reasonable.

## 5. ACKNOWLEDGEMENTS

The authors would like to acknowledge the considerable help provided by A. Sabel'nikov, one of the authors of the Maslov et al. paper [8]. He provided details of their experiment that were not published and answered other questions to help the authors develop a Fluka model of their experiment.

**Appendix 1. Comparison of Fluka Calculations to Experimental Results of Maslov et al.**

A. Sabel'nikov[1], one of the authors of the Maslov et al. paper [8], has provided a detailed description of the geometry used in the experiment that was used to measure the yield of $^{225}$Ac from an electron beam irradiation at 24 MeV. This is shown in Figure A-1.1. The electron beam exits a 0.6 mm-thick stainless-steel window and travels 45 mm through air, collimated by an 8-mm diameter by 25-mm long copper collimator. The target material is captured in the aluminum target holder. The converter target is 2 mm of tungsten.

---

[1] A. Sabel'nikov, one of the authors of the Maslov et al. [8] paper, provided additional experimental details that were used by the authors to produce the Fluka calculations that compared the experimental and calculated yield of $^{226}$Ra(γ,n)$^{225}$Ra.

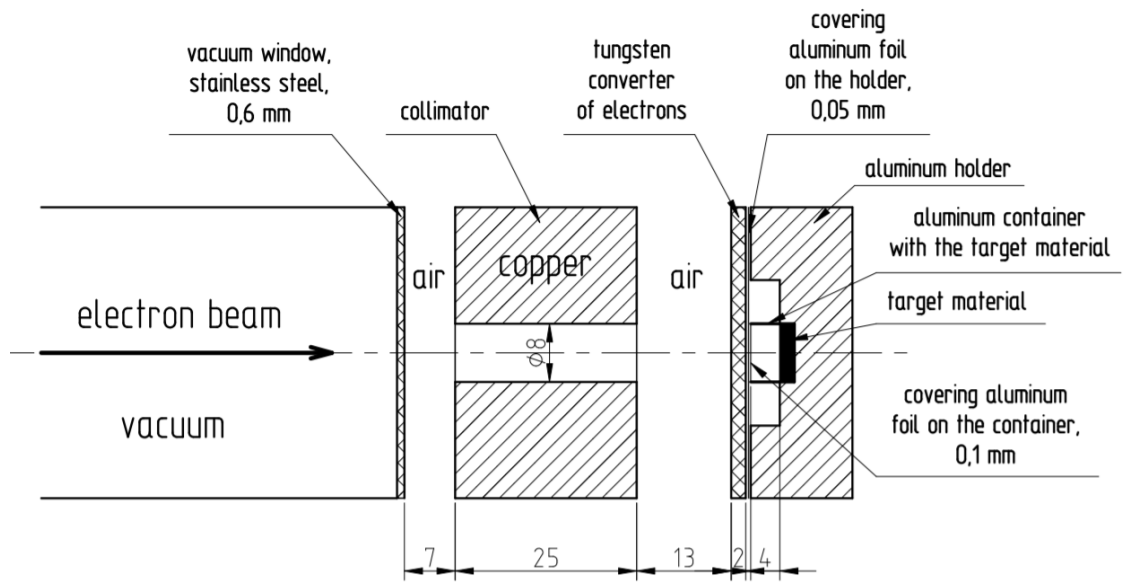

Figure A-1.1. Experimental details of the target used by Maslov et al. [8] to irradiate radium and lead with 24 MeV electrons.

Fluka has been used to calculate the results of the Maslov et al. [8] experiment. Table A-1.1 shows the results of those calculations for several electron-beam diameters from 2 mm (FWHM) to 8 mm (FWHM) for both Gaussian and pencil beam profiles. The experimental beam profile is unknown but is likely Gaussian, between 2 and 6 mm. The effects produced by the changes in beam profile are modest. The Monte Carlo calculations were carried out with a 250 mg target to get reasonable statistics. The extra attenuation in a target of that thickness compared to the thin target used by Maslov et al. [8] should be small. The notes below Table A.2.1 give the details of each column of the calculations.

Table A-2.1. Fluka calculations of the $^{225}$Ac activity of the Maslov et al. [8] experiment for several electron-beam profiles.

| Beam[1] | Ra-225/e[2] | Ra-225[3] Bq/g | Ac-225[4] Bq/g | Ac-225[5] Bq/h-µA-mg | e trans/e[6] | Ac-225[7] (corrected) | Ratio[8] |
|---|---|---|---|---|---|---|---|
| 2 mm FWHM | 1.336 E-5 | 2.901E+8 | 1.289E+8 | 286.5 | 0.693 | 413.4 | 0.75 |
| 4 mm FWHM | 1.235E-5 | 2.683E+8 | 1.192E+8 | 265.0 | 0.621 | 426.8 | 0.78 |
| 6 mm FWHM | 9.988E-6 | 2.169E+8 | 9.642E+7 | 214.3 | 0.485 | 441.9 | 0.80 |
| 8 mm FWHM | 7.596E-6 | 1.650E+8 | 7.332E+7 | 162.9 | 0.357 | 456.3 | 0.83 |
| | | | | | | | |
| 2 mm, pencil | 1.371E-5 | 2.978E+8 | 1.324E+8 | 294.2 | 0.707 | 416.1 | 0.76 |
| 8 mm, pencil | 1.205E-5 | 2.617E+8 | 1.163E+8 | 258.5 | 0.594 | 435.0 | 0.79 |

1: Beam: beam cross section hitting the window.

2: Ra-225/e is the number of Ra-225 atoms produced in the 250 mg target per electron incident on the beam window.

3: Ra-225 (Bq/g) makes use of their irradiation parameters of 30 hours at 15 µA.

4: Ac-225 (Bq/g) is after 18 days, as specified in the Maslov et al. [8] paper.

5: Ac-225 (Bq/(h-µA-mg) is obtained by converting to the units used by Maslov et al. [8].

6: e trans/e: Because of the window thickness and the length of the collimator, a lot of the primary beam is lost in the collimator. This is the fraction of electrons that strike the converter, equivalent to the measured current.

7: Ac-225 (corrected) is the activity expected when expressed in terms of the number of electrons incident on the converter.

8: Ratio: Model results compared to Maslov et al. [8] measurement.

Table A-2.2 shows the results calculated by Fluka for a target of lead chloride of 2.008 grams of natural lead, the same as the larger target used in the Maslov et al. [8] experiment.

Table A-2.2. Fluka calculations of the $^{225}$Ac activity of the Maslov et al. [8] experiment for several electron-beam profiles.

| Beam[1] | Pb-203/e[2] | Pb-203[3] Bq/g | e trans/e[4] | Pb-203[5] corr | 5-day[6] decay | Pb-203[7] Bq/(µA-h-mg) | Ratio, 5 days[8] After EOB MC/meas |
|---|---|---|---|---|---|---|---|
| 2 mm FWHM | 1.837E-6 | 1.910E+5 | 0.697 | 2.742E+5 | 5.516E+4 | 22.1 | 0.79 |
| 4 mm FWHM | 1.684E-6 | 1.751E+5 | 0.624 | 2.808E+5 | 5.649E+4 | 22.6 | 0.81 |
| 6 mm FWHM | 1.344E-6 | 1.398E+5 | 0.487 | 2.870E+5 | 5.775E+4 | 23.1 | 0.82 |
| 8 mm FWHM | 1.025E-6 | 1.066E+5 | 0.359 | 2.972E+5 | 5.980E+4 | 23.9 | 0.85 |
|  |  |  |  |  |  |  |  |
| 2 mm, pencil | 1.881E-6 | 1.956E+5 | 0.71 | 2.752E+5 | 5.537E+4 | 22.2 | 0.79 |
| 8 mm, pencil | 1.699E-6 | 1.767E+5 | 0.60 | 2.959E+5 | 5.954E+4 | 23.8 | 0.85 |

1: Beam: beam cross section hitting the window

2: Pb-203/e is the number of Pb-203 atoms produced in the 2.008 g natural lead as part of a PbCl$_2$ target, per electron incident on the beam window.

3: Pb-203 (Bq/g): Pb-203 activity, per gram of natural lead at EOB. The irradiation conditions were 15 µA for 10 minutes, as specified in the Maslov et al. [8] paper.

4: e trans/e: fraction of electrons striking the converter.

5: Pb-203 (corrected): Pb-203 activity based on electrons striking the converter.

6: 5-day decay: Maslov et al. results were given at 5 days after EOB.

7: Pb-203 (Bq/(h-µA-mg): Result in units given by Maslov et al. [8].

8: MC/meas: Model results compared to Maslov et al. [8] measurement.

The measurement was not a cross section measurement but a demonstration of feasibility and may have some systematic effects. These should be present in both measurements. Both calculations produce results that are about 80% of the experimental results, a good agreement. The consistency between the two results suggests that the cross section of the reaction $^{226}$Ra($\gamma$,n)$^{225}$Ra used in the Fluka code is reasonable.